# Resilient by Design: Simulating Street Network Disruptions across Every Urban Area in the World[1]

Geoff Boeing and Jaehyun Ha
University of Southern California

**Abstract.** Street networks allow people and goods to move through cities, but they are vulnerable to disasters like floods, earthquakes, and terrorist attacks. Well-planned network design can make a city more resilient and robust to such disruptions, but we still know little about worldwide patterns of vulnerability, or worldwide empirical relationships between specific design characteristics and resilience. This study quantifies and measures the vulnerability of the street networks of every urban area in the world then models the relationships between vulnerability and street network design characteristics. To do so, we simulate over 2.4 billion trips across more than 8,000 urban areas in 178 countries, while also simulating network disruption events representing floods, earthquakes, and targeted attacks. We find that disrupting high-centrality nodes severely impacts network function. All else equal, networks with higher connectivity, fewer chokepoints, or less circuity are less vulnerable to disruption's impacts. This study thus contributes a new global understanding of network design and vulnerability to the literature. We argue that these design characteristics offer high leverage points for street network resilience and robustness that planners should emphasize when designing or retrofitting urban networks.

## 1. Introduction

Street networks structure the flow of people and goods through the city. Disruptions to this critical infrastructure sometimes interrupt these flows and cause widespread harm. Heavy rains or sea level rise can flood streets, making many trips impossible. Earthquakes can sever important roadways or block them with debris, complicating evacuations and rescues. Terrorists can target chokepoints such as bridges, tunnels, or important intersections to disconnect parts of the city. Such disruptions can be devastating, but urban street networks are not all equally vulnerable to these catastrophes. Networks designed with more connectivity and redundancy and fewer chokepoints may be more resilient (i.e., adapting and recovering from a disruption over time) and robust (i.e., absorbing or resisting a disruption's impacts in the first place), thus allowing the continued flow of people and goods. Designing such street networks is critical to urban planning for performant and sustainable infrastructure. Urban planners require a better understanding of the relationships between these characteristics and network design—particularly in under-studied but rapidly developing parts of the world—for better evidence-informed planning.

This study addresses this need. It simulates over 2.4 billion trips across more than 8,000 urban areas in 178 countries—covering every urban area in the world larger than a small town. We model these urban area street networks using OpenStreetMap (OSM) data constrained to urban area boundaries defined by the Global Human Settlement Layer. These simulations allow us to answer two questions. First, how vulnerable are street networks around the world to these different types of disruption—in other words, which cities or world regions tend to be more robust or resilient and why? Second, what is the *ceteris paribus* relationship between resilience/robustness and various geometric and topological street network design characteristics?

We find that disrupting high-centrality network nodes leads to particularly severe impacts on network resilience. All else equal, street networks with higher connectivity, fewer chokepoints, or less circuity exhibit less vulnerability to various kinds of disruptions and disasters. This study contributes a new global understanding of worldwide relationships between street network design and street network vulnerability important for both sustainability science and transportation policy and practice. We argue that these design characteristics offer high leverage points for improving network resilience and robustness. Urban planners should in turn emphasize these characteristics when designing new or retrofitting old networks to create more sustainable and reliable urban infrastructure.

The remainder of this article is organized as follows. In the next section, we review recent research into street networks and disambiguate the related terms of resilience, robustness, and efficiency. Then we explain our data and analysis methods, including trip simulations, disruption event simulations, and regression models. Next we present our findings on these disruptions' impacts in different parts of the world and on networks of different design characteristics. Finally, we discuss street network design and vulnerability in the context of real-world urban planning before concluding.

## 2. Background

Street networks play a fundamental role in organizing cities. Their design and structure reflect the legacies of planning paradigms, signal future development, and shape urban transport flows (Amaral and Cunha, 2020; Boeing, 2020; Buhl et al., 2006; Sharifi, 2019). To analyze their form and function, the scientific literature has recently seen strides measuring street networks' geometric and topological features using network science principles (Strano et al., 2013; Shang et al., 2020; Boeing, 2022). This literature usually abstracts street networks as mathematical graphs, where nodes are defined as street intersections and dead-ends and edges are defined as the street segments that link them (e.g., Barthélemy and Flammini, 2008; Porta et al., 2006). That is, a graph consists of a set of nodes connected to each other by a set of edges. Researchers use these graph models to examine street network performance measures such as connectivity, efficiency, and resilience (Boeing, 2020; Buhl et al., 2006; Cavallaro et al., 2014).

One stream of this literature evaluates how different kinds of street networks perform when things go wrong. Unexpected disruptions can impact network performance and several recent studies have explored network vulnerability in the face of such events (e.g., Jenelius and Mattsson, 2012; Miller-Hooks et al., 2012; Sun et al., 2020). Vulnerability is a multifaceted, complex, and fuzzy term intertwined with resilience, reliability, robustness, recovery, redundancy, flexibility, and rapidity (Blockley et al., 2012; Faturechi and Miller-Hooks, 2015; Pan et al., 2021; Wan et al., 2018). Typically, low-vulnerability networks have three characteristics: 1) a low probability of disruptive events, 2) an ability to resist and absorb the effects of such events (often called "robustness"), and 3) the ability to adapt and recover from adverse situations over time (often called "resilience") (Adger, 2000; Gallopin, 2006; Gonçalves and Ribeiro, 2020; Gu et al., 2020; Sun et al., 2020; Westrum, 2012).

Researchers often assess the resilience of street networks by comparing their functionality before and after disruptive events (Dong et al., 2019; Mattsson and Jenelius, 2015; Miller-Hooks et al., 2012; Zhang et al., 2015). One indicator of functionality is origin-destination (OD) connectivity, which measures the probability of OD pairs becoming disconnected due to a disruption (Bell, 2000; Iida and Wakabayashi, 1989; Miller-Hooks et al., 2012; Morelli and Cunha, 2021). Real-world travel demand and capacity constraint data, which are often unavailable, may provide even more realistic measures (Mattsson and Jenelius, 2015). Some researchers study connectivity loss by quantifying the size of the largest connected component after disruptions (Akbarzadeh et al., 2019) and others measure resilience as the loss of accessibility to grocery stores, emergency services, and critical amenities (Alabbad et al., 2021; Dong et al., 2019; Martín et al., 2021; Tsang and Scott, 2020; Wiśniewski et al., 2020; Du et al., 2023). Shang et al. (2022) measure street network robustness by quantifying trips' cost increased based on assumptions of user equilibrium and system optimization.

A central pillar of this body of research explores disruptive events' effects on street networks by simulating node or edge removal (De Montis et al., 2019; Jenelius and Mattsson, 2012; Tang et al., 2020; Yang et al., 2020). This usually follows one of three approaches. First, nodes or edges may be removed deterministically by importance (Akbarzadeh et al., 2019; Du et al., 2023; Duan and Lu, 2014; Martín et al., 2021). This simulates low-entropy disruptions such as targeted terrorist attacks, traffic congestion in key locations, or social conflicts (Mattsson and Jenelius, 2015; Pan et al., 2021). Second, nodes or edges can be removed at random to simulate the impacts of high-entropy spatially distributed processes (Martín et al., 2021; Testa et al., 2015). Third, nodes or edges can be removed based on elevation to simulate disasters such as the flooding of low-lying areas (Kermanshah and Derrible, 2016; Morelli and Cunha, 2021).

Most real-world networks contain critical nodes essential to their overall functionality (Batty, 2013; Jenelius et al., 2006; Novak and Sullivan, 2014; Taylor et al., 2006; Ermagun et al., 2023). Although various methods exist for identifying such important nodes (Gauthier et al., 2018; Scott et al., 2006), betweenness centrality is a common indicator. A node's betweenness centrality represents the (possibly normalized and/or weighted) count of all possible shortest paths in a network that utilize that node (Barthélemy, 2004; Newman, 2005). Akbarzadeh et al. (2019) investigate the impact on network resilience of disrupting nodes with high betweenness centrality, considering different weights. A node or edge with high betweenness centrality in a street network suggests the city's heavy dependence on this element for efficient flows, which can pose a threat to network resilience (Sharifi, 2019). Accordingly, such critical nodes represent high-risk failure points and must be carefully maintained to preserve network functionality (Aydin et al., 2019; Sharifi, 2019).

To varying degrees, street networks are susceptible to disasters. These disruptive events' impacts depend on the geographical location and extent of the disaster (Tsang and Scott, 2020). For example, the highways around New York exhibit high betweenness centralities and suffer flood risks, resulting in low resilience during flooding events (Kermanshah and Derrible, 2017). Ortega (2020) emphasizes the importance of considering the potential future impact of climate change on the most critical network elements. Similarly, Kermanshah and Derrible (2016) demonstrate that the impact of earthquakes on street network resilience is greater in Los Angeles compared to San Francisco, primarily due to the location of critical network elements and the risk of such events.

Some studies suggest that strategically targeting high-centrality nodes in an attack tends to have a larger impact on the network than a random attack (De Montis et al., 2019; Martín et al., 2021). Santos et al. (2021) find that network efficiency declines twice as much in a targeted attack on nodes with high betweenness centrality compared to a random disruption. Similarly, Martín et al. (2021) find that the greatest decrease in network functionality follows the removal of its most critical elements, whereas random removal leads to the least decrease and removing high flood-risk nodes causes a moderate decrease. Ermagun et al. (2023) explore uncertainty in transit network vulnerability by comparing the simulated results of disruptions on the most and least critical links.

Resilient cities need street networks designed to accommodate some amount of failure and to limit connectivity loss due to a disruption (Akbarzadeh et al., 2019; Gonçalves and Ribeiro, 2020; Mattsson and Jenelius, 2015). Gridded networks with small blocks may be more resilient to disruptions than dendritic networks as they have a more-even distribution of centrality values, shorter street segments, more intersections, and higher connectivity (Sharifi, 2019). The average node degree represents an important topological measure of the redundancy and interconnectedness of street networks (Santos et al., 2021; Shang et al., 2020; Strano et al., 2013; Sun et al., 2020). A comparative analysis of Japanese cities by Santos et al. (2021) finds that the average node degree is positively associated with street network robustness. Other studies apply measures such as the network diameter, cyclicity, and link-node ratio to understand the topological determinants of network vulnerability (Ermagun et al., 2023; Shang et al., 2020; Zhang et al., 2015).

This body of literature has advanced the field's understanding of network vulnerability, but also has important limitations. Most of these studies tend to be case studies or focus on specific regions, limiting their generalizability and contribution to universal urban theory. Less research has been done at scale and much remains unknown about the relative resilience and robustness of street networks worldwide. Such knowledge could help unlock a more comprehensive understanding of the geometric and topological features of street network design that contribute to vulnerability, allowing urban planners to build more resilient cities.

## 3. Methods

Our study addresses this need by answering two intertwined questions. First, how vulnerable are street networks worldwide to disruptions from spatially random events, flooding (i.e., elevation-based disruption), or targeted attacks on important nodes? Second, what characteristics are associated with networks that are more robust and efficient against those attacks—in other words, what are the characteristics of the least vulnerable networks? To answer these questions, we model the street networks of every urban area in the world and simulate billions of trips across them. As discussed earlier, resilience refers to the system's ability to recover from a disruption over time and robustness is the system's ability to resist the disruption in the first place. In this study, we operationalize robustness as how many trips become impossible after a disruption event (i.e., the extent to which the network failed to resist the disruption). We also operationalize a measure of efficiency as an indicator of resilience—that is, once we account for robustness, how similarly does the network perform compared to its original performance? This quantifies how far it is from recovering its original performance following the disruption.

### 3.1. Data Sources

This study includes measures of network topology, network geometry, and urban covariates. It examines infrastructure connection and disconnection, not traffic or flows, because our research question concerns trip-level failures rather than traffic impacts to other links in the network cascading from one link's failure. This study uses two main data sources. First, to derive urban area boundaries and socioeconomic covariates around the world, we use the Global Human Settlement Layer (GHSL) Urban Centres Database (UCD) version R2019A[1] (Florczyk et al., 2019). Second, we use the publicly available network models from Boeing (2022) of drivable street networks[2] and related indicators[3], derived from OSM data using the OSMnx Python package and constrained to each UCD urban area's boundary. We retain the strongly connected component (that is, the largest set of network nodes that are fully connected inbound and outbound to each other) of each network model to ensure full OD matrix solvability. Our analysis covers 8,005 urban areas out of the 8,914 in the UCD, as we discard small towns with fewer than 100 network nodes.

**Table 1.** Variables' descriptions, units, and data sources.

| Variable | Description |
|---|---|
| Robustness | Percentage of solvable OD pairs remaining after network disruption event. Derived from OSM data and our simulation analysis. |
| Efficiency | Ratio of the mean reciprocal shortest path distance across all OD pairs ("after the disruption" divided by "before the disruption") expressed as a percentage. Derived from OSM data and our simulation analysis. |
| Avg node deg | Measure of connectedness: average node degree (number of edges per node) of the network. From OSM data via Boeing (2022). |
| Circuity | Ratio of network's street lengths to straight-line distances. From OSM data via Boeing (2022). |
| Chokepoint score | Difference between the max and mean node centrality values, in units of standard deviations. Derived from OSM data and our simulation analysis. |
| Pop density | 1000s of persons per $km^2$ of urban area. From UCD. |
| Pct open space | Percentage of open space in urban area. From UCD. |
| Built up area | 1000s of $km^2$ of built-up urban area. From UCD. |
| Hilliness | Measure of urban area topography: standard deviation of network's node elevations. From OSM and elevation raster data via Boeing (2022). |
| Intersect count | 1000s of intersections. From OSM data via Boeing (2022). |
| Intersect density | 1000s of intersections per $km^2$ of urban area. From OSM, UCD data via Boeing (2022). |
| Flood risk | Important nodes' tendency to concentrate in low-lying areas: correlation coefficient of nodes' centrality and elevation. Derived from OSM and elevation raster data via Boeing (2022) and our analysis. |
| Importance | Relative importance of nodes removed by disruption event: ratio of mean centrality of removed nodes to mean centrality of not-removed nodes. Derived from OSM data and our simulation analysis. |
| Africa | World region dummy for Africa. From UCD. |
| Asia (ex-China) | World region dummy for Asia excluding China. From UCD. |
| China | World region dummy for China. From UCD. |
| Europe | World region dummy for Europe. From UCD. |
| Lat. Am. Carib. | World region dummy for Latin America/Caribbean. From UCD. |
| N. America | World region dummy for Northern America. From UCD. |
| Oceania | World region dummy for Oceania. From UCD. |

Table 1 summarizes our study's variables. We use the average node degree (a measure of network "connectedness"), network circuity, and intersection density from Boeing (2022). We also develop a novel "chokepoint" score, $C$, by calculating the number of standard deviations between each network's maximum and mean node betweenness centrality (henceforth, "centrality"), $B$, for each urban area. Extreme relative dependence on a few nodes (such as a bridge linking two otherwise unconnected parts of a city) suggests points of high fragility. The indicator $C$ is calculated as shown in Equation 1:

$$C = \frac{max(B) - mean(B)}{s.d.(B)} \qquad (1)$$

Next, we calculate the standard deviation of the node elevations for each urban area as a "hilliness" indicator of topography. Finally, we adopt additional urban area-level covariates from the UCD including population density, percent open space, and built-up area. These are described in detail by their creators (Corbane et al., 2017; Freire et al., 2016; Florczyk et al., 2019), but, to summarize, the population data are generated by downscaling spatial census data and the built-up area is identified using information collected from various satellite platforms.

### 3.2. Trip and Disruption Simulations

To measure network resilience and robustness, we simulate trips across each network before and after various network disruptions, using a high-performance computing cluster. First, we randomly sample 10,000 OD pairs for each urban area. As with any such stochastic simulation, the sample size represents a fundamental trade-off between estimate precision and computational time complexity. Our sample size shrinks the standard errors sufficiently for meaningful estimation across all sizes of urban areas. Larger sample sizes offer no precision benefits in terms of the results' interpretation, while increasing the time complexity.

Next we simulate these trips, using Dijkstra's shortest path algorithm weighted by edge length for route solving, before and after network disruptions of different types and magnitudes. The first network disruption type is based on elevation: we eliminate the lowest-lying nodes in the network to approximate a flooding event. The second type is based on importance: we eliminate the nodes with the highest centralities (weighted by distance) in the network to roughly represent a targeted attack on the most important infrastructure, such as in warfare or a terrorist attack. The third type is random: we eliminate a uniform random sample of network nodes, representing high-entropy spatially distributed events like vehicle collisions or earthquakes. Note that these randomized trips do not represent an empirical spatial distribution of real-world trips. Rather, this generates an unweighted characterization of the physical network instead of a weighted characterization accounting for traffic utilization of the physical network. Also note that we avoid within-city inference regarding random disruptions due to the stochasticity of one-time node sampling for this disruption type.

We simulate each of these three disruption types at 0% (the original, undisrupted network), 1%, 2%... up through 10% of the network's nodes eliminated. All told, we simulate 10,000 trips for each urban area 31 times—that is, 3 disruption types × 10 different amounts of nodes eliminated, plus 1 additional simulation with the original undisrupted network as a baseline.

The results of these simulations allow us to operationalize this study's two indicators of network robustness and efficiency. We measure robustness as the percentage of solvable OD pairs that remain after each network disruption, and efficiency as the mean reciprocal shortest path distance across all OD pairs—"after the disruption" divided by "before the disruption"—expressed as a percentage. This robustness indicator, $R$, is calculated as shown in Equation 2 where $J$ is the total number of OD pairs in an urban area, and $J_{p,h}$ is the number of solvable OD pairs after removing $p$ percent of nodes with disruption type $h$:

$$R = \frac{J_{p,h}}{J} \times 100 \qquad (2)$$

The efficiency indicator adapts Latora and Marchiori's (2001) indicator, which they defined as the mean of the reciprocal of shortest path between all node pairs. Our approach follows Santos et al.'s (2021) street network application of this indicator but uses our sampled OD pairs and trip distance instead of time due to inconsistent, unreliable, and unavailable data on travel speeds for network segments around the world. First we calculate the network's "original" efficiency, $E$, as shown in Equation 3 where $J$ is the total number of OD pairs in an urban area and $D$ represents the network distance of each OD pair:

$$E = \frac{1}{J} \sum \frac{1}{D} \qquad (3)$$

Then we calculate the disrupted network's post-event efficiency, $E_{p,h}$, the same way after removing $p$ percent of nodes with disruption type $h$. Finally, we calculate our efficiency indicator (i.e., of the network's "retained" efficiency after disruption), $E^*$, as shown in Equation 4:

$$E^* = \frac{E_{p,h}}{E} \times 100 \qquad (4)$$

### 3.3. Regression Analysis

Finally, we model both our robustness indicator and our efficiency indicator as response variables in a series of six regression models of the form specified in Equation 5:

$$y = \beta_0 + \beta_1 X + \epsilon \qquad (5)$$

where $y$ is the response (either robustness, $R$, or efficiency, $E^*$, after removing 10% of nodes), $\beta_0$ is the intercept, $\beta_1$ is a vector of coefficients to be estimated, $X$ is a matrix of $n$ observations on $k$ predictors, and $\varepsilon$ is the error term. Models I, II, and III predict the robustness indicator, $R$, as the response following 10% centrality-based, elevation-based, and random disruptions respectively. Models IV, V, and VI predict the efficiency indicator, $E^*$, as the response following 10% centrality-based, elevation-based, and random disruptions respectively. We estimate these models' parameters using ordinary least squares. Then we re-estimate each model as a standardized regression to derive beta coefficients expressing magnitudes in units of standard deviations.

The predictors in $X$ include three street network design features of particular interest: the average node degree (i.e., network connectedness), circuity, and chokepoint score. Collectively these measure the redundancy and efficiency of the underlying network, which theoretically may help it resist and recover from disruption. We control for population density, percent open space, built-up area, intersection count (i.e., network size) and density, and hilliness. To specify a model that is linear in parameters, we log-transform the chokepoint score and the built-up area for a better linear fit.

We also develop two additional novel controls. The first controls for a proxy of "flood risk" via the correlation coefficient between node centrality and elevation: low values indicate that low-lying nodes have outsized importance, while high values indicate the opposite. The second controls for a measure of the importance of the nodes we removed for each disruption type, calculated as the mean centrality of removed nodes divided by the mean centrality of not-removed nodes. Finally, we use six world geographical regions, as classified by the United Nations (UNDESA, 2018), as dummy variables to control for region-level differences, omitting Africa as the base class. We treat China separately from the rest of Asia in our analysis due to its enormous number of urban areas (which accounts for 35% and 18% of all the urban areas in Asia and the world, respectively) and documented limitations in Chinese OSM data quality (Zheng and Zheng, 2014). Barrington-Leigh and Millard-Ball (2017; 2019) estimate the coverage and completeness of OSM street network data and find that it is 83% complete with exceptions noted in China.

## 4. Findings

Table 2 lists the mean values of our variables worldwide and per world region. Our three predictors of particular interest are the average node degree, circuity, and chokepoint score. On average, China and Latin America/Caribbean have the highest average node degrees, and Oceania and Asia (excluding China) have the lowest. Northern America and Oceania have the most circuitous street networks on average, whereas China and Latin America/Caribbean have the least. Northern America and Oceania also have the highest average chokepoint scores, and China has the lowest.

### 4.1. Impact of Disruption on Robustness

Figure 1 illustrates the impact of street network disruption on the robustness indicator by visualizing its distribution across each region's urban areas at each 1 percentage point increment of disruption extent. The mean robustness worldwide decreases to 83.1% of the original when the 5% lowest-elevation nodes are eliminated, and it drops to 69.4% when 10% are eliminated. Randomly eliminating 5% of nodes decreases the mean robustness to 80.9%, and it drops to 60.1% when 10% are eliminated. However, targeted centrality-based attacks have the greatest impact. Mean robustness decreases to 89.2% when eliminating just 1% of nodes. Eliminating the 5% highest-centrality nodes decreases mean robustness to 60.5%, and it drops to 33.9% when 10% are eliminated.

Across disruption types and world regions, centrality-based disruptions tend to have the greatest impact on network robustness, followed by random and elevation-based disruptions respectively—though the difference between the latter two is smaller. For example, in Northern America, a 10% centrality-based disruption decreases mean robustness to 23.4%, whereas elevation-based and random disruptions decrease it to just 64.0% and 59.4%, respectively. In Europe, centrality-based, elevation-based, and random disruptions (again affecting 10% of nodes) decrease mean robustness to 24.0%, 65.4%, and 48.0%, respectively. In China, the differences in impact between disruption types tends to be smaller, for reasons discussed in the subsequent discussion section.

**Figure 1.** Robustness indicator results per region, colored by disruption type: y-axis shows indicator value (percent) and x-axis shows disruption extent (percent of nodes removed). Box plots show indicator's distribution across region's urban areas and trend lines track mean value across disruption extents.

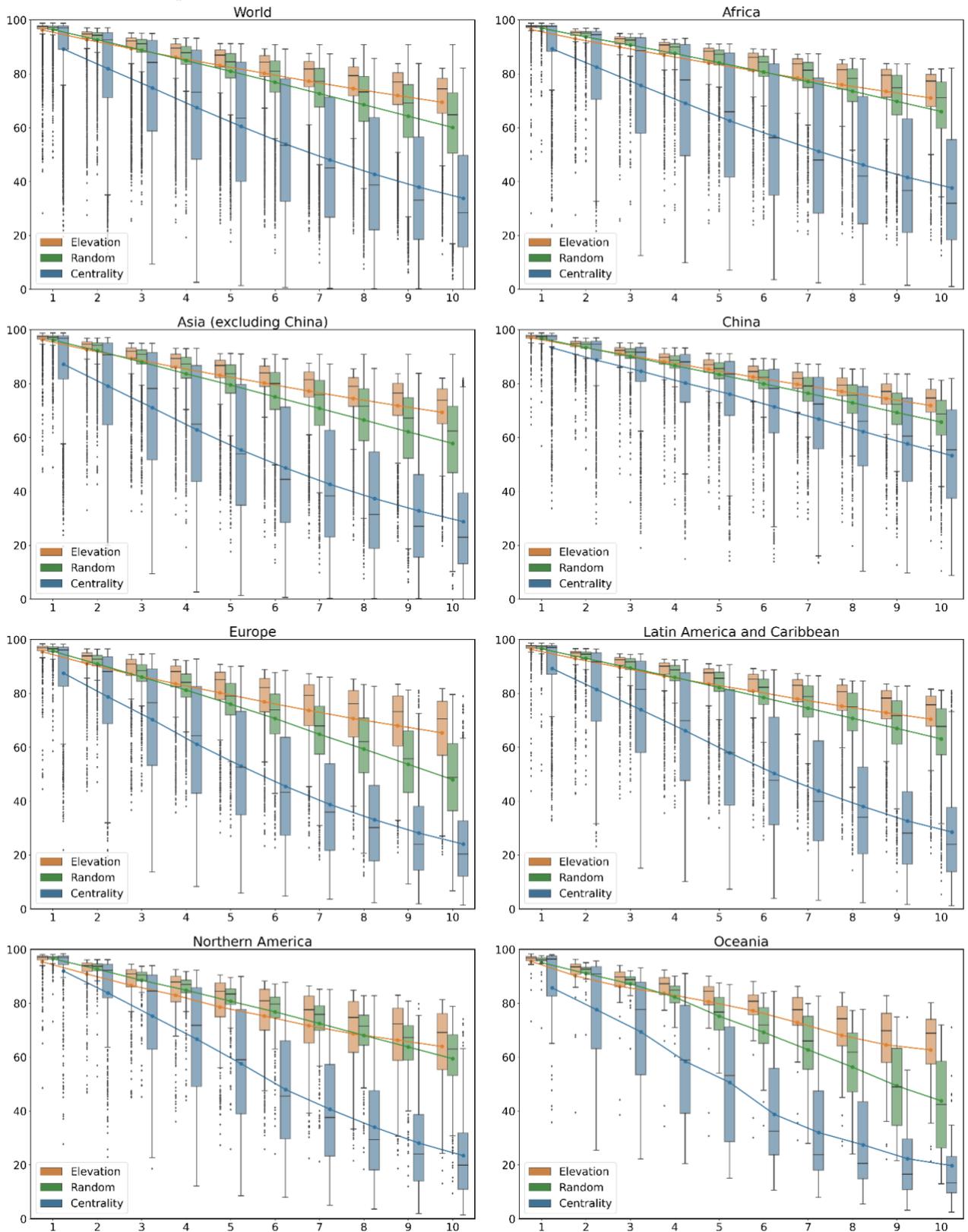

**Figure 2.** Efficiency indicator results per region, colored by disruption type: y-axis shows indicator value (percent) and x-axis shows disruption extent (percent of nodes removed). Box plots show indicator's distribution across region's urban areas and trend lines track mean value across disruption extents.

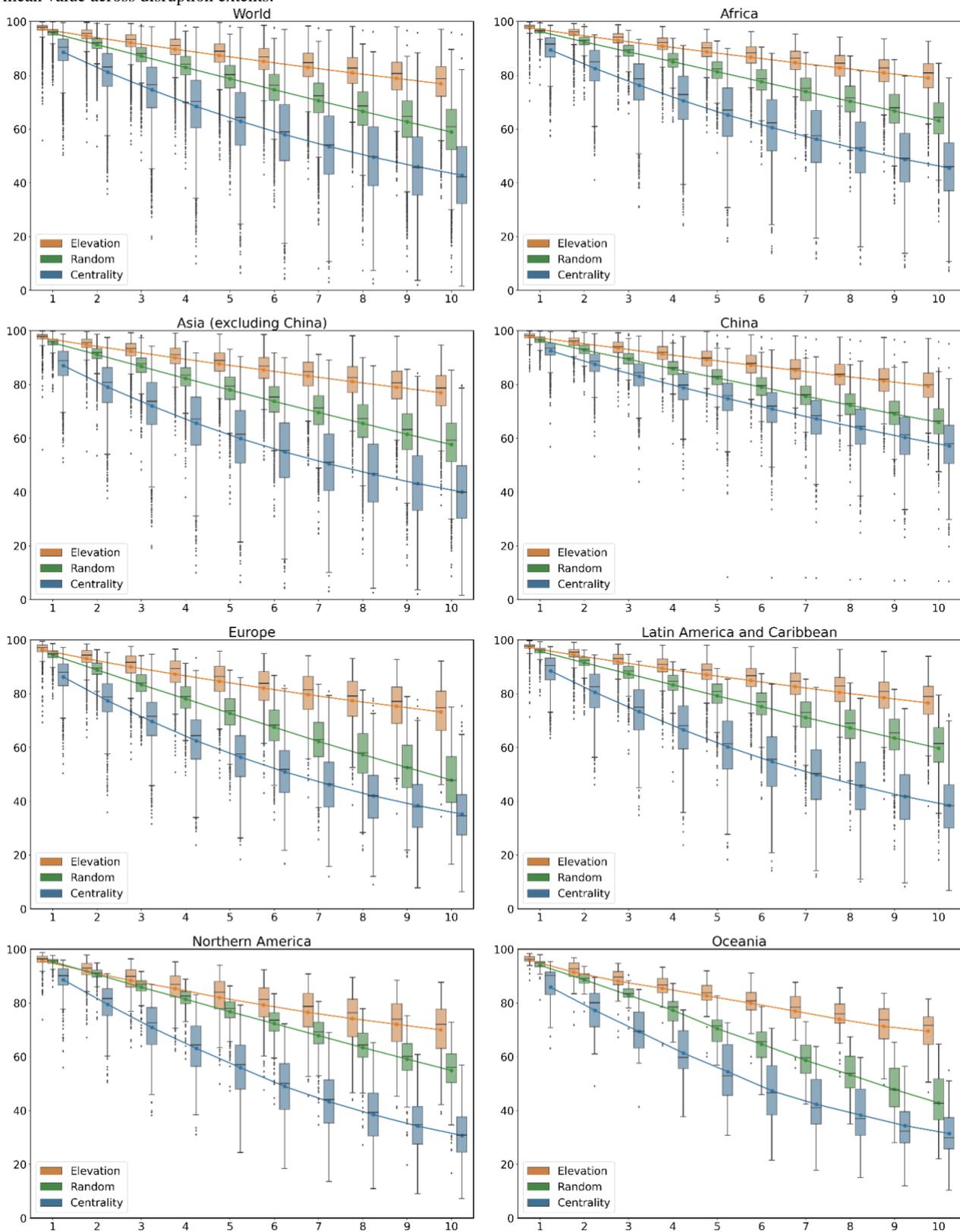

**Table 2.** Mean values of variables per region. All variables/units are as listed in Table 1. Robustness, efficiency, and importance values measure outcomes of simulated centrality-based, elevation-based, or random disruption events affecting 10% of nodes.

|  | World | Africa | Asia (ex-China) | China | Europe | Lat. Am. Carib. | N. America | Oceania |
|---|---|---|---|---|---|---|---|---|
| Robustness | | | | | | | | |
|     Centrality-based | 33.90 | 37.65 | 28.92 | 53.27 | 24.03 | 28.53 | 23.41 | 19.70 |
|     Elevation-based | 69.42 | 71.04 | 69.34 | 71.80 | 65.36 | 70.45 | 63.95 | 62.68 |
|     Random | 60.10 | 66.00 | 57.82 | 65.79 | 47.98 | 63.09 | 59.41 | 43.75 |
| Efficiency | | | | | | | | |
|     Centrality-based | 42.87 | 45.49 | 40.20 | 57.19 | 35.20 | 38.50 | 30.72 | 31.44 |
|     Elevation-based | 76.88 | 79.01 | 77.01 | 79.39 | 73.20 | 76.57 | 69.99 | 69.52 |
|     Random | 58.94 | 63.11 | 57.72 | 66.01 | 47.87 | 59.71 | 54.89 | 42.83 |
| Avg node degree | 2.82 | 2.86 | 2.70 | 2.93 | 2.72 | 3.00 | 2.85 | 2.67 |
| Circuity | 1.05 | 1.05 | 1.06 | 1.04 | 1.06 | 1.04 | 1.07 | 1.07 |
| Chokepoint score | 6.70 | 6.73 | 6.80 | 4.56 | 7.38 | 7.78 | 9.35 | 7.97 |
| Pop density | 23.00 | 35.76 | 33.45 | 13.33 | 6.23 | 17.05 | 2.39 | 6.63 |
| Pct open space | 67.25 | 73.96 | 73.84 | 59.28 | 60.11 | 63.96 | 56.81 | 48.50 |
| Built up area | 0.04 | 0.02 | 0.03 | 0.03 | 0.04 | 0.03 | 0.18 | 0.11 |
| Hilliness | 17.16 | 16.76 | 14.61 | 13.23 | 21.08 | 25.00 | 19.41 | 21.65 |
| Intersect count | 4.63 | 3.60 | 5.49 | 0.93 | 4.54 | 5.92 | 12.83 | 11.78 |
| Intersect density | 0.20 | 0.31 | 0.24 | 0.03 | 0.12 | 0.30 | 0.07 | 0.12 |
| Flood risk | -0.01 | 0.02 | 0.01 | -0.03 | -0.05 | -0.03 | -0.05 | -0.06 |
| Importance | | | | | | | | |
|     Centrality-based | 9.91 | 9.69 | 11.26 | 5.43 | 11.43 | 9.78 | 14.52 | 12.02 |
|     Elevation-based | 0.74 | 0.66 | 0.66 | 0.72 | 0.94 | 0.78 | 1.02 | 1.02 |
|     Random | 1.00 | 1.00 | 1.00 | 1.01 | 1.00 | 1.01 | 1.00 | 0.97 |
| N | 8005 | 1423 | 2659 | 1454 | 1049 | 1007 | 372 | 41 |

**Table 3.** Regression model parameter estimates, standard errors, and beta coefficients. All variables/units are as listed in Table 1. * and bold indicate estimate significance at $p<0.05$.

| | Model I robustness / centrality | | | Model II robustness / elevation | | | Model III robustness / random | | | Model IV efficiency / centrality | | | Model V efficiency / elevation | | | Model VI efficiency / random | | |
|---|---|---|---|---|---|---|---|---|---|---|---|---|---|---|---|---|---|---|
| | coef | s.e. | beta | coef | s.e. | beta | coef | s.e. | beta | coef | s.e. | beta | coef | s.e. | beta | coef | s.e. | beta |
| Constant | -13.409 | 9.298 | 0.045 | **67.263*** | 6.202 | -0.022 | **56.007*** | 7.489 | 0.219 | 4.696 | 4.679 | 0.021 | **83.825*** | 3.870 | 0.062 | **58.461*** | 4.323 | 0.196 |
| Avg node degree | **33.577*** | 0.964 | 0.373 | **10.123*** | 0.608 | 0.196 | **32.840*** | 0.730 | 0.499 | **20.375*** | 0.485 | 0.346 | **6.165*** | 0.379 | 0.173 | **22.600*** | 0.421 | 0.490 |
| Circuity | -20.602 | 7.277 | -0.027 | **-10.653*** | 4.889 | -0.025 | **-62.035*** | 5.853 | -0.113 | **-10.956*** | 3.662 | -0.022 | **-13.412*** | 3.051 | -0.045 | **-43.482*** | 3.378 | -0.113 |
| Chokepoint score (log) | -15.627 | 0.790 | -0.280 | **-2.541*** | 0.510 | -0.079 | **-3.272*** | 0.612 | -0.080 | **-12.599*** | 0.397 | -0.345 | **-3.155*** | 0.318 | -0.143 | **-5.685*** | 0.353 | -0.199 |
| Pop density | **0.066*** | 0.007 | 0.089 | **0.018*** | 0.005 | 0.042 | **0.035*** | 0.006 | 0.065 | **0.033*** | 0.004 | 0.068 | **0.011*** | 0.003 | 0.037 | **0.032*** | 0.003 | 0.085 |
| Pct open space | -0.011 | 0.014 | -0.008 | **-0.090*** | 0.009 | -0.123 | **-0.076*** | 0.011 | -0.081 | **0.044*** | 0.007 | 0.053 | **-0.040*** | 0.006 | -0.078 | -0.002 | 0.006 | -0.003 |
| Built up area (log) | **-2.441*** | 0.267 | -0.144 | **-0.969*** | 0.173 | -0.100 | **-1.786*** | 0.207 | -0.144 | **-3.638*** | 0.134 | -0.328 | **-1.100*** | 0.108 | -0.164 | **-2.207*** | 0.120 | -0.254 |
| Hilliness | **0.029*** | 0.010 | 0.024 | **0.131*** | 0.007 | 0.188 | -0.000 | 0.008 | -0.000 | -0.005 | 0.005 | -0.006 | **0.128*** | 0.004 | 0.266 | **-0.013*** | 0.005 | -0.021 |
| Intersect count | **0.123*** | 0.011 | 0.107 | **0.014*** | 0.007 | 0.021 | **0.046*** | 0.008 | 0.054 | **0.028*** | 0.006 | 0.037 | 0.000 | 0.004 | 0.000 | **0.032*** | 0.005 | 0.054 |
| Intersect density | **-11.351*** | 1.134 | -0.108 | 0.040 | 0.755 | 0.001 | **-6.488*** | 0.907 | -0.084 | **-14.907*** | 0.571 | -0.216 | **-1.940*** | 0.471 | -0.047 | **-10.499*** | 0.523 | -0.195 |
| Flood risk | **9.470*** | 1.403 | 0.055 | **-3.279*** | 1.098 | -0.033 | **6.477*** | 1.128 | 0.052 | **2.463*** | 0.706 | 0.022 | 0.948 | 0.685 | 0.014 | **2.870*** | 0.651 | 0.033 |
| Importance | **-0.673*** | 0.042 | -0.206 | **-16.050*** | 0.322 | -0.547 | **-16.320*** | 0.779 | -0.174 | **-0.165*** | 0.021 | -0.077 | **-11.155*** | 0.201 | -0.550 | **-13.264*** | 0.449 | -0.202 |
| Asia (ex-China) | **-2.113*** | 0.516 | -0.095 | 0.603 | 0.346 | 0.047 | **-2.038*** | 0.415 | -0.126 | **-1.593*** | 0.260 | -0.110 | -0.371 | 0.216 | -0.042 | **-1.325*** | 0.240 | -0.117 |
| China | **6.029*** | 0.708 | 0.271 | 0.435 | 0.472 | 0.034 | **-3.442*** | 0.567 | -0.212 | **5.793*** | 0.356 | 0.399 | 0.106 | 0.295 | 0.012 | -0.533 | 0.328 | -0.047 |
| Europe | **-1.898*** | 0.694 | -0.085 | 0.632 | 0.464 | 0.050 | **-10.388*** | 0.556 | -0.641 | **-1.234*** | 0.349 | -0.085 | **-0.690*** | 0.289 | -0.078 | **-8.417*** | 0.321 | -0.740 |
| Lat. Am. Carib. | **-9.031*** | 0.645 | -0.406 | **-0.894*** | 0.432 | -0.070 | **-6.246*** | 0.518 | -0.385 | **-4.641*** | 0.325 | -0.319 | **-2.033*** | 0.269 | -0.231 | **-3.906*** | 0.299 | -0.344 |
| N. America | -0.033 | 0.982 | -0.001 | 0.719 | 0.659 | 0.056 | -0.858 | 0.790 | -0.053 | **-1.825*** | 0.494 | -0.126 | **-1.810*** | 0.411 | -0.206 | **-0.997*** | 0.456 | -0.088 |
| Oceania | -2.740 | 2.413 | -0.123 | -0.814 | 1.615 | -0.064 | **-12.721*** | 1.939 | -0.784 | -1.004 | 1.214 | -0.069 | **-2.876*** | 1.008 | -0.327 | **-10.983*** | 1.119 | -0.966 |
| $R^2$ | 0.548 | | | 0.383 | | | 0.451 | | | 0.733 | | | 0.496 | | | 0.628 | | |
| $n$ | 8005 | | | 8005 | | | 8005 | | | 8005 | | | 8005 | | | 8005 | | |

Across world regions, Northern America and Oceania suffer the greatest robustness impacts from centrality-based disruptions: their mean robustness values decrease to 23.4% and 19.7%, respectively, after eliminating the 10% highest-centrality nodes. In contrast, China (53.3%) and Africa (37.7%) suffer the least such impact. Northern America (64.0%) and Oceania (62.7%) again suffer the greatest robustness impacts from elevation-based 10% disruptions, whereas China (71.8%) and Africa (71.0%) suffer the least. Finally, Oceania (43.8%) and Europe (48.0%) suffer the greatest robustness impacts from random 10% disruptions, whereas China (65.8%) and Africa (66.0%) again suffer the least.

**4.2. Impact of Disruption on Efficiency**

Figure 2 illustrates the impact of street network disruption on the efficiency indicator by visualizing its distribution across each region's urban areas at each 1 percentage point increment of disruption extent. The mean efficiency worldwide decreases to 87.4% of the original when the 5% lowest-elevation nodes are eliminated, and it drops to 76.9% when 10% are eliminated. Randomly eliminating 5% of nodes decreases the mean efficiency to 78.7%, and it drops to 58.9% when 10% are eliminated. As with the robustness indicator, centrality-based disruptions have the greatest impact on efficiency. Mean efficiency decreases to 88.6% when eliminating just 1% of nodes. Eliminating the 5% highest-centrality nodes decreases mean efficiency to 62.9%, and it drops to 42.9% when 10% are eliminated.

As we previously saw with the robustness indicator, across disruption types and world regions, centrality-based disruptions tend to have the greatest impact on network efficiency, followed by random and elevation-based disruptions respectively. For example, in Northern America, a 10% disruption based on centrality decreases mean efficiency to 30.7%, whereas elevation-based and random disruptions decrease it to just 70.0% and 54.9%, respectively. In Europe, centrality-based, elevation-based, and random disruptions (again affecting 10% of nodes) decrease mean efficiency to 35.2%, 73.2%, and 47.9%, respectively. The efficiency indicator value gaps between disruption types grow as disruption extents increase.

Across world regions, Northern America and Oceania suffer the greatest efficiency impacts from centrality-based disruptions: their mean efficiency values decrease to 30.7% and 31.4%, respectively, after eliminating the 10% highest-centrality nodes. In contrast, China (57.2%) and Africa (45.5%) suffer the least such impact. Northern America (70.0%) and Oceania (69.5%) again suffer the greatest efficiency impacts from elevation-based 10% disruptions, whereas China (79.4%) and Africa (79.0%) suffer the least. Finally, Oceania (42.8%) and Europe (47.9%) suffer the greatest efficiency impacts from random 10% disruptions, whereas China (66.0%) and Africa (63.1%) again suffer the least.

**4.3. Regression Results**

Table 3 presents the results of our six regression models of the relationship between street network characteristics and the robustness and efficiency indicators across all three disruption types. The key variables of interest include average node degree, the chokepoint score, and circuity. We find that all three have statistically significant ($p<0.05$) coefficient estimates across these models. In particular, the average node degree and chokepoint score tend to have relatively large standardized beta coefficients, suggesting relatively large effect sizes. These models explain between 38-73% of the variation in our two indicators. Across the two indicators, the models predicting efficiency have higher coefficients of determination ($R^2$) on average, and across the three disruption types the models of centrality-based disruption events have higher $R^2$ values on average.

Urban street networks with higher average node degrees remain significantly more robust and efficient across centrality-based, elevation-based, and random disruption events. In other words, networks with more connected nodes on average are less vulnerable to these three types of disasters. All else equal, a 1 unit increase in average node degree is associated with a 33.6, 10.1, and 32.8 percentage point increase in robustness on average following the simulated centrality-based, elevation-based, and random disruptions, respectively. Similarly, a 1 unit increase in average node degree is associated with a 20.4, 6.2, and 22.6 percentage point increase in efficiency across these disruption types. In particular, it has the largest effect size following random disruption events, where a 1 standard deviation increase in average node degree is associated with roughly a 0.4 standard deviation increase in both robustness and efficiency.

Conversely, urban street networks with higher chokepoint scores consistently exhibit significantly lower robustness and efficiency across all three disruption types. In other words, networks that depend heavily on a few nodes are more vulnerable and less resilient to these three types of disasters. All else equal, a 1 percent increase in the chokepoint indicator is associated with a 0.16, 0.03. and 0.03 percentage point decrease in robustness on average following the simulated centrality-based, elevation-based, and random disruptions, respectively. Similarly, a 1 percent increase in the chokepoint indicator is associated with a 0.13, 0.03, and 0.06 percentage point decrease in efficiency across these disruption types. In particular, it has the second largest effect size following centrality-based disruption events, where a 1 standard deviation increase in the log-transformed chokepoint score is associated with a 0.28 and 0.35 standard deviation decrease in robustness and efficiency, respectively.

Finally, urban street networks with higher circuity values consistently exhibit significantly lower robustness and efficiency across all three disruption types, though the effect size is smaller. In other words, more circuitous networks are more vulnerable and less resilient to these three types of disasters. All else equal, a 1 unit increase in circuity is associated with a 20.6, 10.7, and 62.0 percentage point decrease in robustness on average following the simulated centrality-based, elevation-based, and random disruptions, respectively. Similarly, a 1 unit increase in circuity is associated with a 11.0, 13.4, and 43.5 percentage point decrease in efficiency across these disruption types. In particular, it has the largest effect size following random disruption events, where a 1 standard deviation increase in circuity is associated with a 0.1 standard deviation decrease in both robustness and efficiency.

## 5. Discussion

As the world urbanizes, the sustainability and reliability of its urban transport infrastructure becomes ever more important. These infrastructure networks are vulnerable—to varying degrees—to different disasters. Our findings reveal how vulnerable street networks around the world are to different types of disruptions and provide estimates of the relationships between such vulnerability and a basket of network design characteristics.

Returning to our research questions: first, how vulnerable are networks to disruption? We argue that it generally depends on the disruption type and the location. Our findings reveal that attacking nodes with high centrality leads to particularly severe network disconnection and reduced trip efficiency. Random disruptions result in moderate impacts and elevation-based disruptions result in the least. Among world regions, urban areas in Northern America and Oceania were particularly vulnerable. Different world regions have different planning legacies, design paradigms, and cultures reflected in their present day infrastructure and resilience outcomes. Northern America and Oceania exemplify this: these two regions have the highest circuity and chokepoint scores (and Oceania had the lowest average node degrees), and in turn they consistently rank poorly across robustness and efficiency outcomes following disruption events.

Our second research question asked: what network design characteristics are associated with less vulnerable networks? We find consistent and significant relationships between the robustness/efficiency indicators and our three street network design variables of particular interest across all six regression models. All else equal, urban street networks with 1) higher average node degrees, 2) lower chokepoint scores, or 3) lower circuity values exhibit higher robustness and efficiency values following centrality-based, elevation-based, or random disruptions.

To unpack this, we can consider examples of *centrality*-based disruption in four urban areas. Table 4 presents the average node degree and chokepoint score (the two predictors of interest with particularly large effect sizes), and the robustness and efficiency indicators. Wichita and Toulouse exhibit higher robustness and efficiency compared to Salt Lake City and Amsterdam when disrupting 10% of nodes by centrality. Wichita has a higher average node degree and a lower chokepoint score than Salt Lake City, illustrating how connectedness shapes street network vulnerability. Figure 3 shows how Wichita's street network consists primarily of small grids, whereas the Salt Lake City's street network includes numerous dead ends within large blocks, concentrating node importance on the block perimeters. Meanwhile, rivers bisect both Toulouse and Amsterdam, making high-centrality bridges (i.e., chokepoints) critical for connecting the urban area. Figure 3 shows that all of Amsterdam's important bridges were affected by the disruption event (and are distributed throughout the urban area). Conversely, Toulouse has a couple of lower-centrality bridges that remained after the disruption event, allowing trip connections. This illustrates the difference in Amsterdam's and Toulouse's chokepoint scores and the fact that >40% of the OD pairs remain connected in Toulouse, whereas <5% of the OD pairs remain connected in Amsterdam.

In Table 5 and Figure 4, we present another set of four examples, this time of *elevation*-based disruptions. These disruption simulations had minimal impacts on Granada and San Antonio where almost 80% of the OD pairs remained connected. Figure 4 shows that their low-lying nodes are mainly located in peripheral areas, leading to minimal impact on robustness and efficiency. However, Cape Coral's low-lying nodes are scattered across the urban area, and disrupting those nodes disconnects >70% of the simulated OD pairs. Jacksonville illustrates a different case where the low-lying nodes are spatially concentrated, disconnecting the urban area and resulting in low robustness and efficiency values. The location of impacted nodes is essential for understanding street network vulnerability to elevation-based disruption.

Overall and in sum, this study's results make theoretical sense. Higher average node degrees add redundancy to the network through more points of connectivity. This redundancy allows for fallback options when components fail. In contrast, the chokepoint score suggests an over-reliance on critical points that can devastate network function if they fail. Accordingly, the network's average node degree and chokepoint score exhibited high beta coefficients across the regression models. This point implies an important takeaway for urban planners: these particular design characteristics may offer high leverage points for network resilience, because the average node degree and the chokepoint score each independently have large standardized relationships with robustness and efficiency. These findings suggest that urban planners and engineers should emphasize designing new—and retrofitting old—networks with more redundancy and less reliance on chokepoints (when topography and costs allow) to create more robust and resilient cities.

**Table 4.** Selected measures of four urbanized areas' street networks. Robustness and efficiency indicator values represent centrality-based disruption simulations affecting 10% of nodes.

| Urban area | Avg node deg | Chokepoint score | Robustness | Efficiency |
|---|---:|---:|---:|---:|
| Wichita, Kansas, USA | 3.00 | 9.20 | 58.57 | 38.10 |
| Toulouse, France | 2.79 | 8.53 | 45.64 | 36.43 |
| Salt Lake City, Utah, USA | 2.66 | 14.56 | 8.27 | 13.52 |
| Amsterdam, Netherlands | 2.80 | 16.51 | 4.19 | 12.64 |

**Figure 3.** Examples of simulated centrality-based disruption events affecting 10% of nodes. Colored edges are removed by the simulated node disruption. Map scales vary.

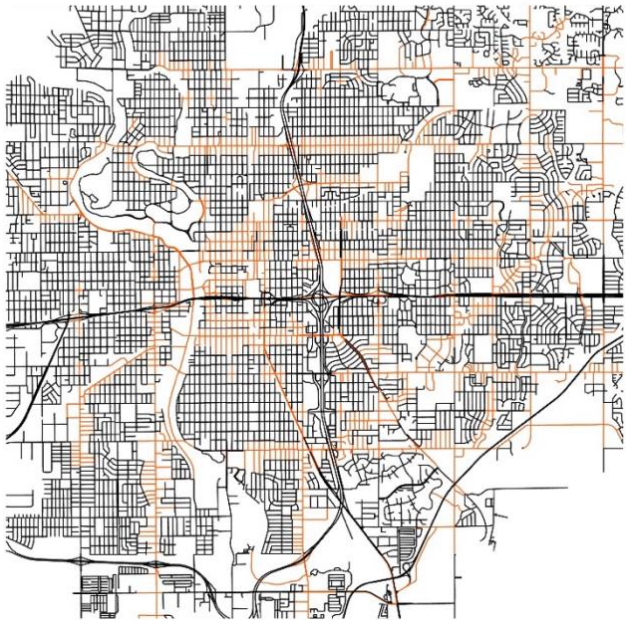
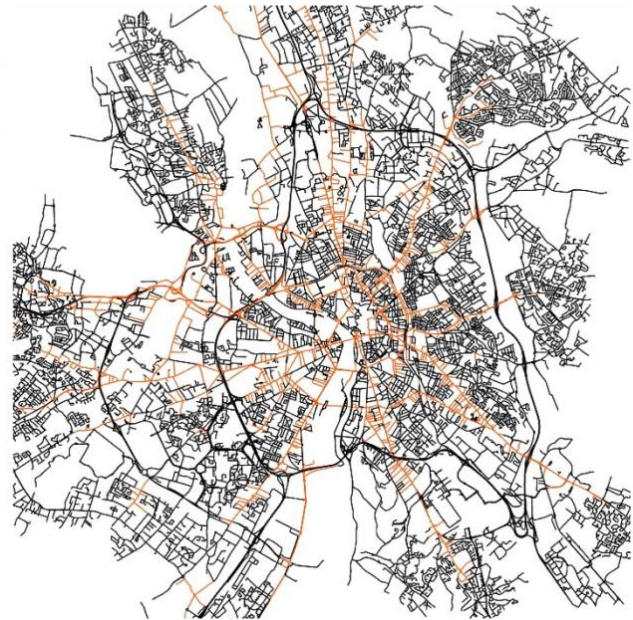
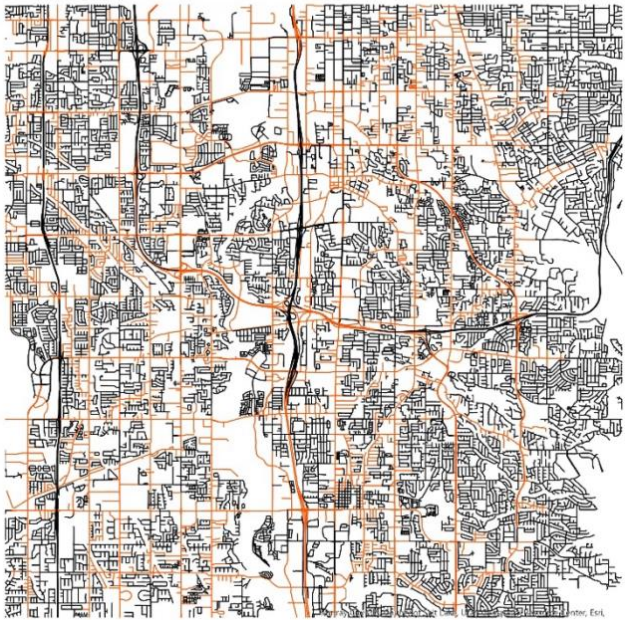
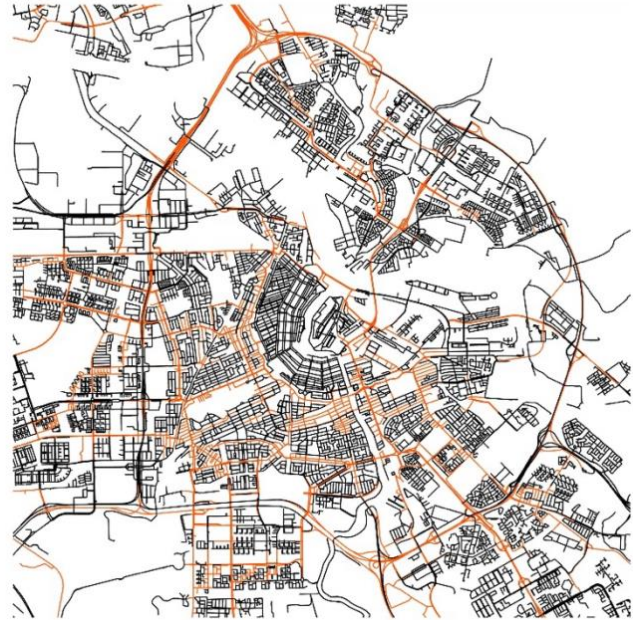

**Table 5.** Selected measures of four urbanized areas' street networks. Robustness and efficiency indicator values represent elevation-based disruption simulations affecting 10% of nodes.

| Urban area | Avg node deg | Chokepoint score | Robustness | Efficiency |
|---:|---:|---:|---:|---:|
| Granada, Spain | 2.91 | 8.68 | 79.77 | 83.15 |
| San Antonio, Texas, USA | 2.90 | 11.07 | 79.71 | 79.86 |
| Cape Coral, Florida, USA | 2.94 | 10.71 | 25.24 | 38.78 |
| Jacksonville, Florida, USA | 2.87 | 15.28 | 24.64 | 42.22 |

**Figure 4.** Examples of simulated elevation-based disruption events affecting 10% of nodes. Colored edges are removed by the simulated node disruption. Map scales vary.

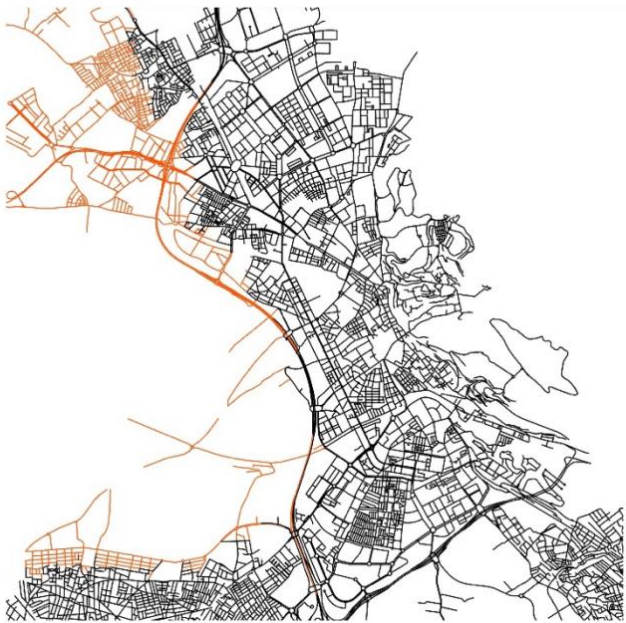
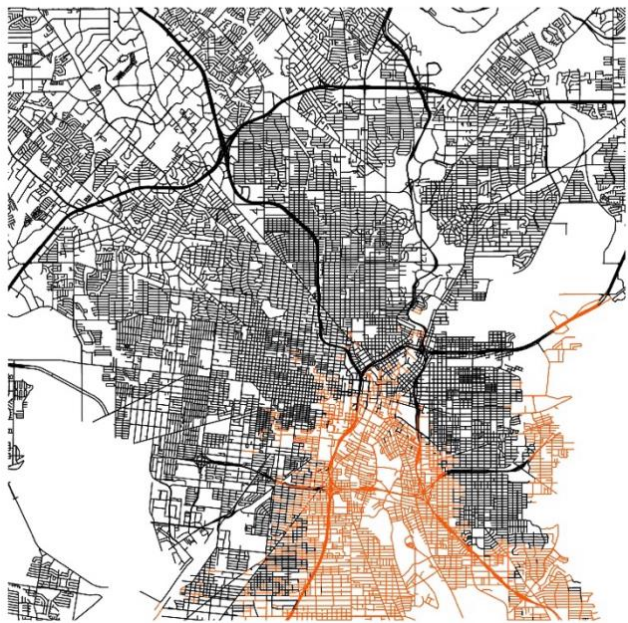
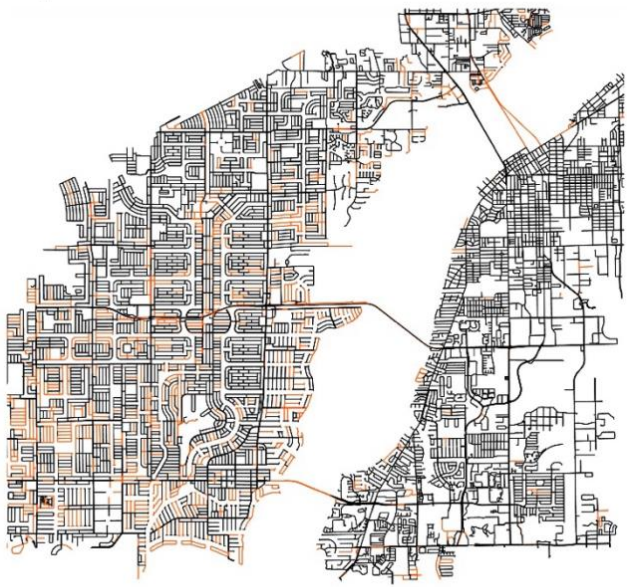
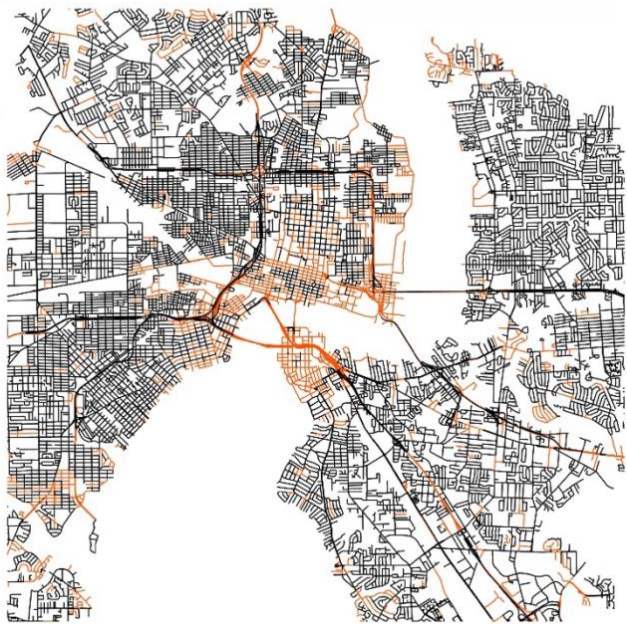

With that in mind, this study's limitations open the door for future research. First, a global study's data needs inherently limit variable selection to globally available and relatively consistent data. Shifting this research to finer spatial scales could supplement these findings with local case studies using bespoke data that may be unavailable globally. For example, this could include the incorporation of travel speeds and local economic conditions. Additional topographical data could help reveal physical constraints to road building. Examining a city's growth over time can reveal how it evolves and why, and in turn how it shapes resilience and robustness.

Second, and related, our randomized OD pairs do not represent empirical trips. On one hand, our unweighted characterization of the physical network does provide a relatively even and complete characterization of the physical structure itself. On the other hand, this is not a weighted characterization that accounts for network utilization through traffic and flows. Empirical trip data do not exist worldwide today, but future work can perform such an analysis locally where the data do exist.

Third, there is unevenness in how urban areas distribute around the planet. For example, less-populated Oceania offers us only 41 urban areas, which limits inference within this region. Related, the trend in China differs from the rest of Asia and the Chinese results may be biased due to aforementioned local OSM data quality issues. For example, the differences across disruption types are relatively small in China, possibly due to incomplete OSM networks affecting our node centrality measurements, which could influence downstream simulation outcomes. Given its rate of urbanization, China in particular presents a pressing case for future research. Nevertheless, our study takes the first steps in providing much-needed worldwide and regional estimates of these critical relationships to advance urban science and the evidence base for more sustainable urban planning.

## 6. Conclusion

Street networks are essential to urban economies and citizens' lives and livelihoods. Disruption to this infrastructure from disasters such as earthquakes, floods, or targeted attacks can cause significant harm. Yet we have lacked a broad understanding of worldwide relationships between street network design and street network resilience. This study addressed this gap. It simulated billions of trips across more than 8,000 urban areas around the world to 1) ask how vulnerable are networks around the world to different disruptions, and 2) what geometric and topological network characteristics are associated with higher or lower vulnerability?

Our findings suggest that networks designed with more connectivity and redundancy are more robust and efficient following a disruption, allowing people and goods to continue to flow. Designing such networks is essential to the mission of building sustainable, resilient cities. We argue that planners require a stronger evidence base, particularly in under-studied and rapidly developing world regions, for designing better transport networks. This study provides a step in that direction and opens the door to ongoing research in this urgently needed area.

## Notes

UCD: http://data.europa.eu/89h/53473144-b88c-44bc-b4a3-4583ed1f547e
[2] Global Urban Street Networks GraphML repository v2: https://doi.org/10.7910/DVN/KA5HJ3
[3] Global Urban Street Networks Indicators repository v2: https://doi.org/10.7910/DVN/ZTFPTB